\documentclass[12pt,preprint]{aastex}

\shorttitle{STAR FORMATION AND AGNS IN QSOS}
\shortauthors{Kimball et al.}

\begin{document}
\bibliographystyle{apj}

\title{The Two-Component Radio Luminosity Function of QSOs: Star Formation and AGN}
\author{Amy E. Kimball\altaffilmark{1,2}}
\author{K. I. Kellermann\altaffilmark{2}}
\author{J. J. Condon\altaffilmark{2}}
\author{\v{Z}eljko Ivezi\'{c}\altaffilmark{3}}
\author{Richard A. Perley\altaffilmark{4}}

\altaffiltext{1}{akimball@nrao.edu}
\altaffiltext{2}{National Radio Astronomy Observatory, 520 Edgemont Rd., Charlottesville, VA 22903, USA}
\altaffiltext{3}{Department of Astronomy, University of Washington, Box 351580,
  Seattle, WA 98195, USA}
\altaffiltext{4}{National Radio Astronomy Observatory, Socorro, NM 87801, USA}

\begin{abstract}
Despite decades of study, it remains unclear whether there are distinct
radio-loud and radio-quiet populations of quasi-stellar objects (QSOs).  Early
studies were limited by inhomogeneous QSO samples, inadequate sensitivity to
probe the radio-quiet population, and degeneracy between redshift and
luminosity for flux-density-limited samples.  Our new 6 GHz EVLA observations
allow us for the first time to obtain nearly complete (97\%) radio detections
in a volume-limited color-selected sample of 179 QSOs more luminous than $M_i =
-23$ from the Sloan Digital Sky Survey (SDSS) Data Release Seven in the narrow
redshift range $0.2 < z < 0.3$.  The dramatic improvement in radio continuum
sensitivity made possible with the new EVLA allows us, in 35 minutes of
integration, to detect sources as faint as 20 $\mu$Jy, or
$\log[L_\mathrm{6\,GHz}({\rm W\,Hz^{-1}})]\approx21.5$ at $z = 0.25$, well
below the radio luminosity, $\log[L_6({\rm W\,Hz}^{-1})]\approx22.5$, that
separates star-forming galaxies from radio-loud active galactic nuclei (AGNs)
driven by accretion onto a super-massive black hole.  We calculate the radio
luminosity function (RLF) for these QSOs using three constraints: (a) EVLA 6
GHz observations for $\log[L_6({\rm W\,Hz^{-1}})]<23.5$, (b) NRAO-VLA Sky
Survey (NVSS) observations for $\log[L_6({\rm W\,Hz^{-1}})]>23.5$, and (c) the
total number of SDSS QSOs in our volume-limited sample.  We show that the RLF
can be explained as a superposition of two populations, dominated by AGNs at
the bright end and star formation in the QSO host galaxies at the faint end. 
\end{abstract}

\keywords{quasars: general --- galaxies: active --- galaxies: starburst}

\section{INTRODUCTION}
\label{sec:intro}

Since the discovery of optically selected QSOs \citep{sandage65}, the
difference between the ``radio-loud" (RLQ) and ``radio-quiet" (RQQ) QSO
populations remains elusive.  There has been a continuing controversy as to
whether the radio luminosity distribution of QSOs is bimodal
\citep[e.g.,][]{kellermannEtal89,i02,i04qso} or is merely broad and smooth
\citep[e.g.,][]{cirasuoloEtal03,lacyEtal10}.  A bimodal distribution suggests
that two distinct physical processes are present, with one process being
significantly more powerful than the other. 

The two primary sources of radio emission from galaxies are (1) accretion onto
super-massive black holes in active galactic nuclei (AGNs) and (2) star
formation.  Radio-AGN emission is due to synchrotron from relativistic plasma
jets, and the associated hotspots and lobes resulting from jet interaction with
the surrounding medium.  Star formation yields free-free emission from HII
regions and synchrotron radiation from relativistic electrons believed to be
accelerated in supernova remnants.  Radio AGNs can be extremely luminous, with
radio emission reaching $\log[L_\mathrm{6\,GHz}({\rm W\,Hz}^{-1})]\approx27$ in
the case of 3C~273 \citep{3Cspectra}, whereas star formation leads to radio
luminosities of $\log[L_\mathrm{6\,GHz}({\rm W\,Hz}^{-1})]\approx21$ for a
galaxy like the Milky Way, or as high as $\log[L_\mathrm{6\,GHz}({\rm
    W\,Hz}^{-1})]\approx23.5$ in the case of a star-bursting galaxy such as Arp
220.  Radio-AGNs dominate the radio luminosity function (RLF) of galaxies
brighter than $\log[L_\mathrm{6\,GHz}({\rm W\,Hz}^{-1})]\approx22.5$ in the
radio, while star-formation in galaxies without an AGN dominates at fainter
luminosities \citep{condonEtal02}. 

The question as to the nature of radio emission in radio-quiet QSOs has not
been previously addressed using a homogeneous and complete QSO sample that is
sensitive to luminosities significantly fainter than
$\log[L_\mathrm{radio}({\rm W\,Hz}^{-1})]=22.5$.  It is possible that some, if
not all, QSOs are hosted in star-forming galaxies, which only become visible at
radio wavelengths when the AGN radio emission is below some given threshold.
In this Letter, we use results from the Expanded Very Large Array (EVLA)
\citep{introPaper} to constrain the radio luminosity function (RLF) of QSOs.
We show that the RLF is consistent with the hypothesis that QSO radio sources
with 6 GHz spectral luminosity $\log[L_6({\rm W\,Hz}^{-1})]>{23}$ are powered
primarily by AGNs, while those with $\log[L_6({\rm W\,Hz}^{-1})]<23$ are
powered primarily by star formation in their host galaxies. 

By taking advantage of recent advances in both radio and optical capabilities,
we have, for the first time, obtained nearly complete radio detections in a
large volume-limited sample of optically  selected QSOs.  In this Letter, we
describe early results of our EVLA observations.  In Section~\ref{sec:data} we
describe the target selection and radio observations.  In
Section~\ref{sec:results} we present our results and interpret them in terms of
the superimposed radio contributions from AGNs and star-forming host galaxies.
Our conclusions are summarized in Section~\ref{sec:conclusions}.

\section{DATA}
\label{sec:data}

\subsection{Target selection}

Our color-selected targets were drawn from the \citet{dr7quasars} quasar
catalog of the seventh data release (DR7) of the Sloan Digital Sky Survey
(SDSS) \citep{dr7}.  We selected nearby QSOs in order to reach a spectral
luminosity limit of $\log[L_6({\rm W\,Hz^{-1}})]\approx21.5$ after 35 minutes
of integration with the 2\,GHz bandwidth available at the EVLA, and in a narrow
redshift range to minimize the effects of evolution.  Our selection criteria
were: 
\begin{enumerate}
\item targeted by the SDSS for spectroscopy using the ``low $z$" color criteria
  (see explanation below) 
\item $M_\mathrm{i} < -23$ ($H_\mathrm{0}=70$ km s$^{-1}$ Mpc$^{-1}$, $\Omega_\Lambda=0.7$)
\item $14 < i < 19$  \citep[apparent magnitude $i$ corrected for extinction
  according to][]{sfd}, which is well above the SDSS completeness limit 
\item $0.2 < z < 0.3$
\item $b > 30^\circ$
\end{enumerate}

We used the SDSS to select a homogeneous sample of SDSS QSOs.  The SDSS
identifies low-redshift QSO candidates for follow-up spectroscopy by their
non-stellar colors in the SDSS $ugri$ color cube.  These sources are labeled in
the SDSS database with the QSO targeting flag ``low z".  QSOs are then selected
from all objects with spectra, and identified by having at least one broad
emission line with FWHM greater than 1000~km~s$^{-1}$ \citep{richards02}. 

The selection criteria result in 179 targets that comprise a volume-limited,
color-selected sample of all $M_\mathrm{i}<-23$ QSOs in $\sim$2.66 steradians
in the redshift range $0.2 < z < 0.3$.  Owing to the color selection criteria,
our sample may miss some QSOs in our search volume that happen to have stellar
colors.  However, it constitutes a fully optically selected sample with no
radio biases.  About one-third of QSOs in our target list were already known
radio emitters found in the Faint Images of the Radio Sky at Twenty cm (FIRST)
\citep{first} survey (peak flux density $S_{\rm p}>1$~mJy at $\nu=1.4\,$GHz)
and/or found in the NRAO---VLA Sky Survey (NVSS) ($S_{\rm p}>2.5$ ~mJy at
$\nu=1.4\,$GHz) \citep{nvss}.

\subsection{Observations}

All of the observations were made with the C configuration of the EVLA using a nominal 
bandwidth of 2 GHz centered on 6 GHz in each of two circular polarizations, which were 
combined to make total intensity images.  Because any source detected in the FIRST or NVSS 
surveys is expected to be at least a few hundred microjanskys at 6 GHz, we observed the QSOs 
identified with FIRST or NVSS radio sources for only three minutes.  To minimize sidelobes in 
the dirty beam we tried to divide these short snapshot observations into two 90~s 
segments widely separated in hour angle, but owing to scheduling constraints about 10\% of 
these FIRST/NVSS sources were observed for only 90~s at one hour angle.  All
such sources were detected in the single 90~s observation.  All other QSOs
were, at first, observed for five minutes.  Most of those that were clearly
detected in the five minute observation were re-observed for another five
minutes at a different hour angle.  The remaining sources were observed for an
additional 30--35 minutes, their observations were combined, and the resulting
images generally reach the expected rms noise levels of 6--8$\mu $Jy per
synthesized beam solid angle ($\sim3\farcs5$ FWHM). 

For the purposes of this Letter, we consider sources to be detected at a
$3\sigma$ confidence level.  This confidence level is appropriate, given that
the QSO positions are known with $\sim0\farcs1$ accuracy from the SDSS
\citep{sdss_astrometry}.  Less than 10\% of the QSOs in our sample have
3--$4\sigma$ detections; the probability than any one of these is false is
$<0.0014$.  Thus the probability that our sample contains even one false
detection is $<2.5\%$.  

\section{AGN and Galaxy Components of QSO Radio Emission}
\label{sec:results}

\begin{figure}
\epsscale{0.6}
\plotone{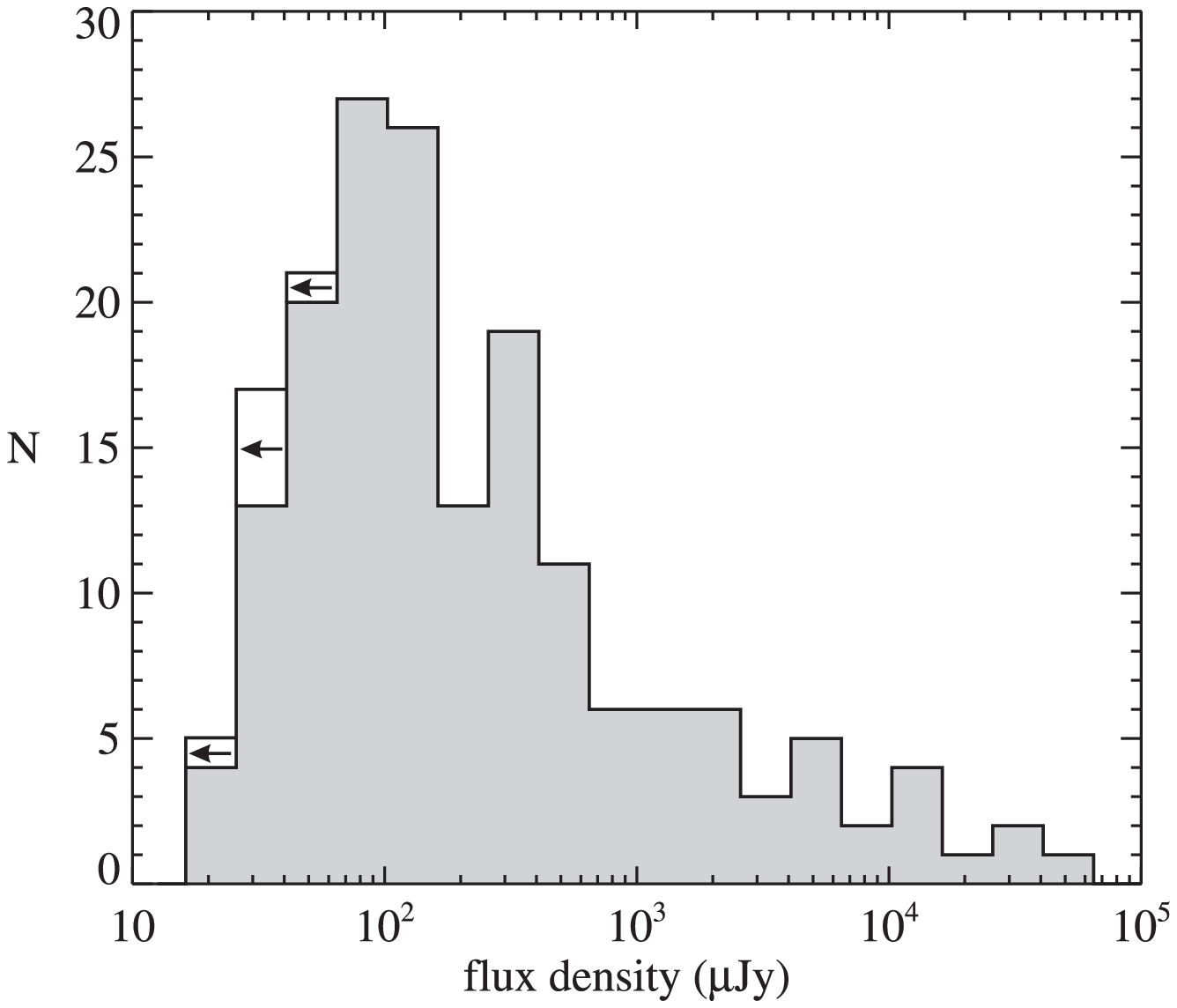}
\plotone{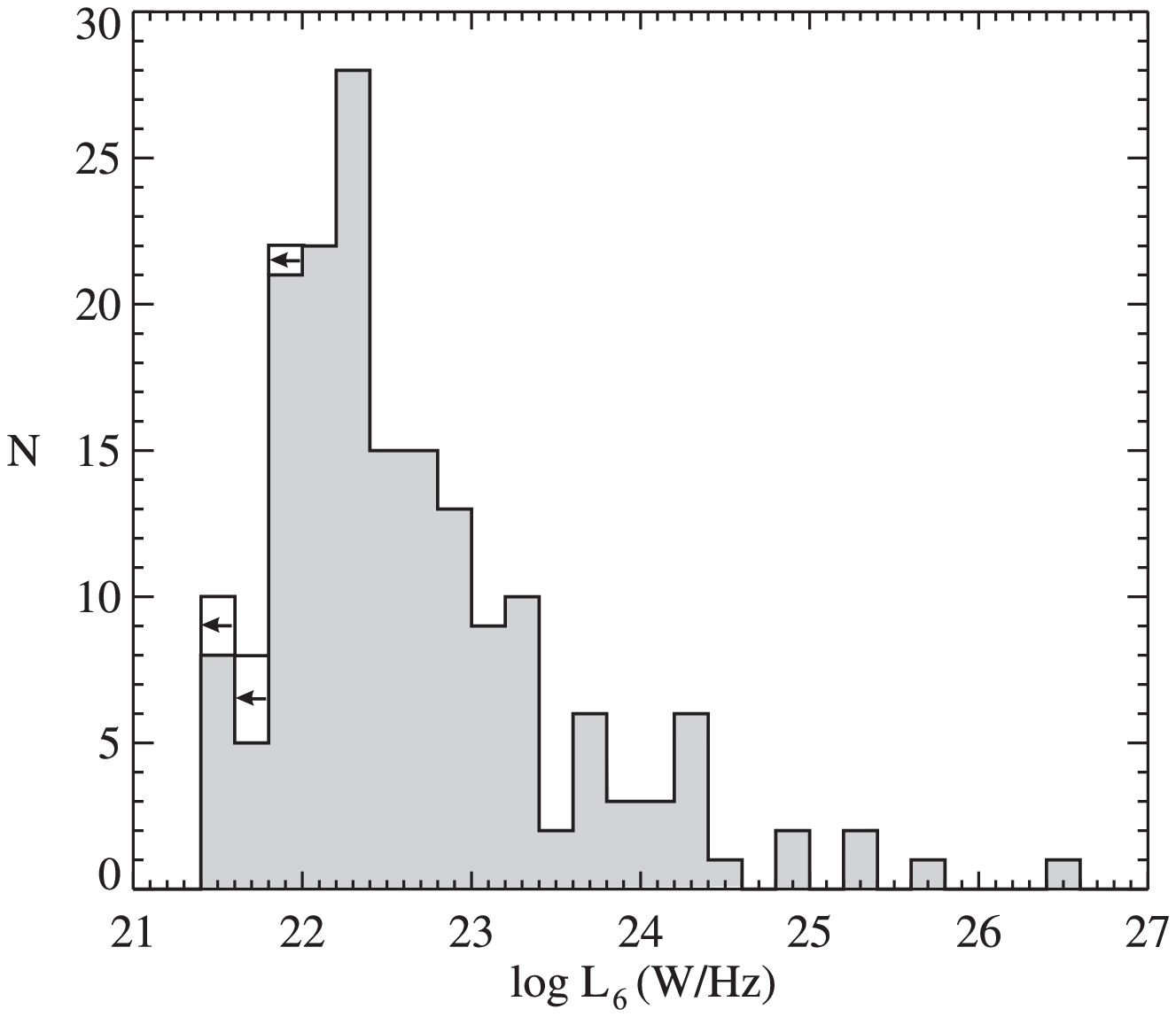}
\caption{Top: Histogram of the distribution of 6~GHz flux densities observed by
  the EVLA in our sample of 179 QSOs.  For six undetected sources, 3$\sigma$
  upper limits are shown (arrows).  Bottom: Intrinsic spectral radio power of
  the 179 QSOs at 6 GHz in the source frame.} 
\end{figure}

In the top panel of Figure~1, we show the distribution of observed 6 GHz flux
densities for the compact (unresolved on $\sim3\farcs5$ scale) radio components
coincident with the 179 QSOs in our color-selected sample.  In the bottom
panel, we show the corresponding spectral luminosity distribution calculated
assuming a spectral index $\alpha = -0.7$ (where $L_\nu \propto \nu^\alpha$;
see discussion below) for the $K$-correction \citep{kcorrect}.  The observed
distributions are clearly peaked around 100 $\mu$Jy and a few times $10^{22}$
W\,Hz$^{-1}$ respectively, which are well above our detection limits.  Only six
QSOs remain undetected.  For one of these, 103421.71+605318.1, there was a
strong 69 mJy source in the field, which degraded the sensitivity limit by a
factor of 2.5.  For the other five undetected QSOs, the rms noise ranges from 7
to 11 $\mu$Jy, somewhat above our nominal limit owing to interference, limited
on-source observing time, or inadequate dynamic range.  We include the
$3\sigma$ upper limits of these six undetected sources in Figure 1. 

For the purposes of this Letter, we assume that all of the QSOs have a spectral
index of $\alpha=-0.7$ between 6 GHz and 1.4 GHz.  A direct measurement of the
true spectral index for a faint FIRST source is difficult, as a significant
fraction of emission can be missed from faint sources owing to ``clean bias"
\citep{nvss}.  The NVSS resolution ($\sim45\arcsec$ beam) is not a good match
to the resolution of our EVLA observations ($\sim3\farcs5$ beam); thus it can
also be difficult to measure accurate spectral indices for extended NVSS
sources in the sample.  The 6~GHz EVLA observations suggest that, except for
the very strongest QSOs, the sources appear unresolved.  Therefore, we can use
NVSS images to estimate the 1.4 GHz flux density of the remaining QSOs, even
though they do not have counterparts in the NVSS catalog. We measured the flux
densities at the optical positions of all QSOs not identified with cataloged
NVSS sources stronger than the catalog limit of 2.4 mJy.  The flux-density
distribution of these QSOs has a median value of 330 +- 30 microJy.  The median
flux density of our 6 GHz EVLA detections is 126~$\mu$Jy, yielding an estimate
of $\alpha=-0.69$ for the average spectral index of the QSO sample.  This value
is typical of star-forming galaxies \citep{condonReview92}.  Using the actual
spectral indices of individual sources would not significantly change Figure~1
nor the analysis discussed in the remainder of this section, as a change in
spectral index of $\Delta\alpha=0.5$ corresponds to a factor of two in
luminosity and a difference of only 0.3 on a $\log$ scale. 

The 6 GHz spectral luminosity function of $0.2<z<0.3$ color-selected QSOs
derived from these observations is shown by the black data points with error
bars in Figure~2.  The luminosity function was calculated using the
$V/V_\mathrm{max}$ method on detected sources \citep{schmidt68}.  The
$V/V_\mathrm{max}$ method statistically determines the RLF using detected
quantities; upper limits are not used in the calculation.  Instead, undetected
sources are accounted for by the normalization of the accessible volume ($V$)
of the detected sources.  Because there are only six undetected QSOs, there
cannot be more than six undetected QSOs in the luminosity range $18.8 <
\log[L_6({\rm W\,Hz}^{-1})] < 21.2$, which fact we indicate by the wide
upper-limit symbol in Figure~2.  Also shown are the 6 GHz luminosity functions
of nearby ($z < 0.05$) galaxies whose radio luminosities are dominated either
by AGNs (solid green curve) or star formation (solid red curve), calculated
from the 1.4 GHz luminosity functions \citep{condonEtal02} using spectral index
$\alpha=-0.7$.  The 6 GHz spectral luminosity function of nearby galaxies whose
radio sources are dominated by AGNs extends to higher radio luminosities than
the luminosity of star-forming galaxies, so AGN-powered sources are more common
above $\log [L_6 ({\rm W\,Hz}^{-1})] \approx 22.5$; star-forming galaxies have
higher space densities than AGNs at lower luminosities.  The nearby
``starburst'' galaxy M82, which is a typical radio-selected star-forming
galaxy, has $\log[L_6({\rm W\,Hz}^{-1})]\approx21.5$.  This value is close to
the $10^{21}$ W Hz$^{-1}$ spectral luminosity of the Milky Way at 6 GHz
\citep{berkhuijsen84}. 

Above $\log[L_6({\rm W\,Hz}^{-1})] \approx 23.5$, the spectral luminosity of
the ultraluminous starburst galaxy Arp 220, it is likely that the QSO radio
emission is primarily produced by AGNs.  To estimate the luminosity function
that would result if all QSO radio emission were powered entirely by AGNs, even
at lower luminosities, we extrapolated the high-luminosity QSO luminosity
function to faint luminosities as shown by the dashed green curve in Figure 2
for $21<\log[L_6(\rm{W\,Hz}^{-1})]<25.5$ using a calculated slope of $-0.30
\pm0.03$.  The slope for this extrapolation is not adequately constrained by
the small number of $\log[L_6(\rm{W\,Hz}^{-1})]>24$ QSOs in our sample (16 of
179).  Instead, we determined the slope from the observed 1.4 GHz QSO
luminosity function, using NVSS \citep{nvss} measurements at the positions of
SDSS QSOs in the redshift range $0.2<z<0.45$.  The lack of observed QSOs with
very high radio luminosities motivates the fall-off at $\log[L_6({\rm
    W\,Hz}^{-1})]>26$, although its exact form is not known and cannot be
determined from our data.  However, this fall-off does not affect the
calculation of the power-law slope below
$\log[L_6(\rm{W\,Hz}^{-1})]\approx25.5$.  The slope is consistent with the EVLA
data points for $\log[L_6({\rm W\,Hz}^{-1})]>24$, as it should be.  The dashed
green curve is shown extrapolated to $\log[L_6({\rm (W Hz)}^{-1})] \approx
21.3$ to emphasize that it falls below the observed luminosity function in this
range; either the AGN luminosity function has a "bump" not observed in the AGN
luminosity function of nearby galaxies or there is an additional energy source
contributing to the radio emission of most QSOs with $21 < \log[L_6({\rm W
    Hz}^{-1})] < 23$.  If the luminosity function of the AGN component alone is
extrapolated to still lower luminosities (as suggested by the dotted green
line), it must fall off around $\log[L_6({\rm W Hz}^{-1})] \sim 19$ lest the
number of AGN exceed the total number of QSOs in our SDSS sample.  The exact
form of the luminosity function and its cutoff indicated by the dotted curve is
unknown but not critical, because it always predicts that most QSOs powered
only by AGN would be weaker than the EVLA detection limit $S \approx
20\,\mu$Jy, or $\log[L_6({\rm W Hz}^{-1})] > 21.3$ if $z > 0.2$. 

\begin{figure}
\epsscale{0.9}
\plotone{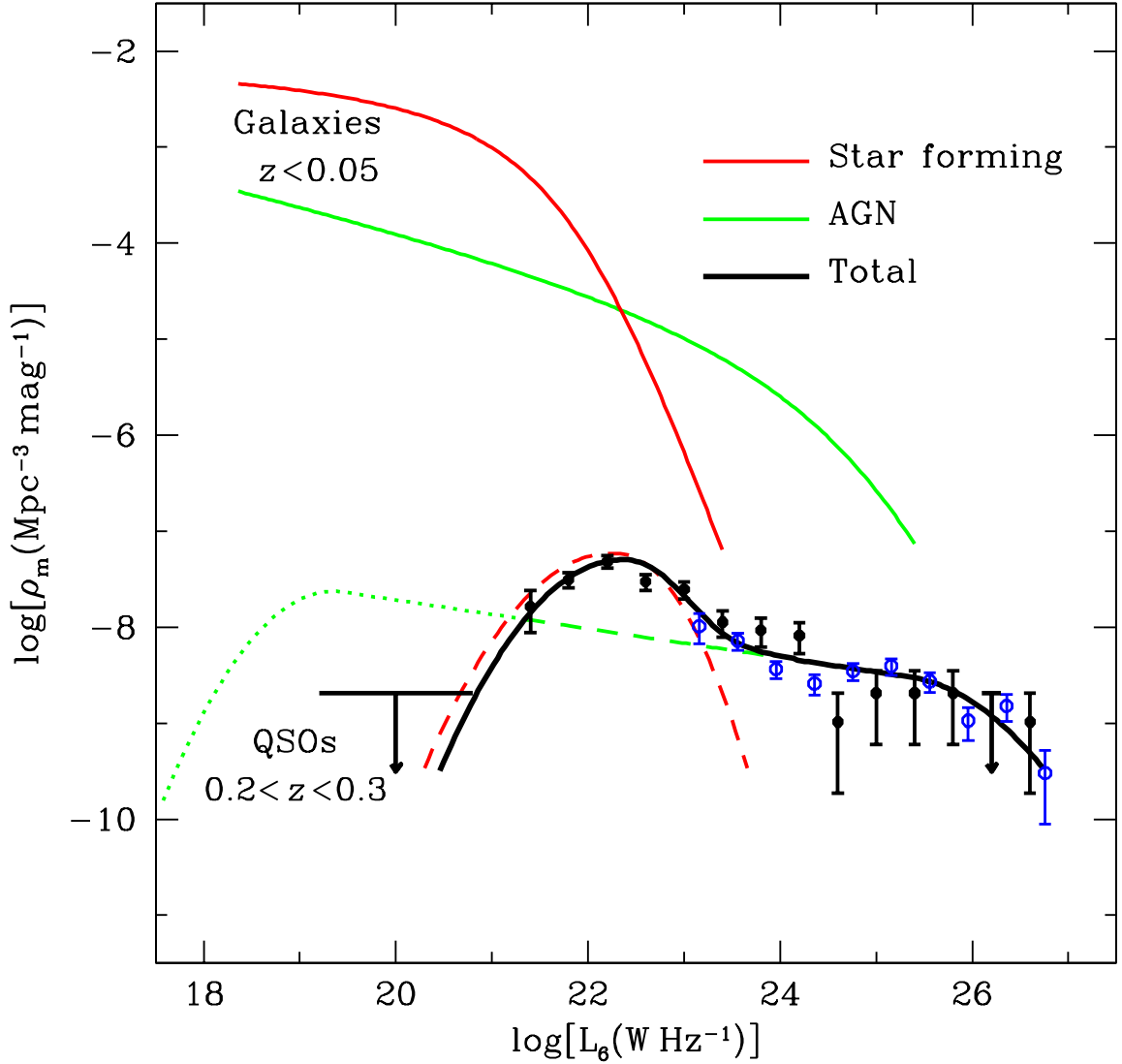}
\caption{\label{fig:rhomfig} 
Our models for the spectral luminosity functions of $0.2 < z < 0.3$ QSOs are
compared to the 1.4 GHz luminosity function of nearby galaxies
\citep{condonEtal02} transformed to 6 GHz with an assumed spectral index of
-0.7. Radio sources powered primarily by star formation are shown with a solid
red curve and those powered by AGNs as a solid green curve.  Solid black points
correspond to our EVLA data; open blue points correspond to NVSS sources.  The
dashed green curve is an extrapolation of the high-radio-luminosity QSOs to low
luminosities, using the slope determined from the NVSS data.  The dashed red
curve represents the spectral luminosity function of QSO hosts that are powered
primarily from star formation, constrained by the EVLA data.  The upper limit
symbol at $\log[L_6({\rm W\,Hz}^{-1})] \sim 20$ represents the six EVLA
non-detections.  While it is not certain that all of the non-detections have
luminosities this low (see Figure 1), this upper limit constrains the most
conservative estimate of the RLF fall-off (see text for details).  The black
curve is the calculated luminosity function for QSOs whose radio sources are
powered by both AGNs and star formation in their host galaxies.  The total area
under the black curve is constrained by the number of SDSS QSOs in the
volume-limited ($0.2 < z < 0.3$) sample.} 
\end{figure}

In order to be consistent with both the EVLA detections of QSOs stronger than
$20\,\mu$Jy and the total number of SDSS QSOs, the QSO RLF {\it must} rise
sharply just below $\log[L_6({\rm W\,Hz}^{-1})] \approx 23.5$ and fall fairly
sharply at lower luminosities.  {\it The total number of SDSS QSOs in this
  luminosity function is known (179 for $0.2<z<0.3$), and limits the integral
  area of the RLF.}  Even if all six non-detections are not in the luminosity
range $18.8 < \log[L_6({\rm W\,Hz}^{-1})] < 21.2$, the upper limit in Figure~2
constrains the most conservative estimate of the RLF fall-off.  If any of the
six non-detections actually have radio luminosity in the range $21.2 <
\log[L_6({\rm W\,Hz}^{-1})] < 22$, the RLF at low luminosities must fall even
more sharply than what is shown in Figure~2. 

We suggest that the rise in the RLF at $\log[L_6({\rm W\,Hz}^{-1})]\approx 
23.5$ is the result of radio emission from the star-forming host galaxies.
The solid red curve in Figure~2 shows the 6 GHz space density of radio sources
powered by all star-forming galaxies in the local universe
\citep{condonEtal02}.  The dashed red curve illustrates one possible form of
the 6 GHz spectral-luminosity function for the host galaxies of $0.2 < z < 0.3$
QSOs, a parabolic fit with fall-off determined independently on either side of
the peak.  The peak of the curve is at $\log[L_6({\rm W\,Hz}^{-1})]=22.2$ and
$\log[\rho_{\rm m}({\rm Mpc^{-3}\,mag^{-1}})]=-7.2$.  The slope of the fall-off
on either side of the peak is not tightly constrained by our data.  The
fall-off slope indicated by of our dashed red curve at $\log[L_6({\rm
    W\,Hz}^{-1})]>22.5$ is consistent with the fall-off slope of the solid red
curve, but with 3\% of the space density.  This result suggests that the space
density of luminous starbursts in QSOs at $0.2<z<0.3$ is approximately 3\% the
space density of local galaxies powered by starbursts.  However, this fraction
may be different if the true RLF distribution of star-forming QSO host galaxies
is not well-represented by the dashed red curve. 

The black curve is the corresponding luminosity function of QSOs whose radio
luminosities are the sum of both the AGN and star-forming galaxy luminosities,
on the assumption that the AGN and star-forming luminosities are statistically
independent.  The spectral-luminosity functions indicated by the dashed red
curve and by the black curve are constrained by the EVLA detections shown as
black data points in Figure~2 and by the total number of SDSS QSOs.  The peak
of the QSO flux-density distribution implied by the black curve is $\langle
\log[S(\mu{\rm Jy})] \rangle \approx 2.1$ at $\nu = 6{\rm ~GHz}$, and the peak
of the the QSO host-galaxy luminosity function should be at $\log[L_6({\rm
    W\,Hz}^{-1})] \approx 22.4$. 

As it must, the RLF of the QSO host galaxies (dashed red curve) lies below the
RLF of {\it all} galaxies powered primarily by star formation (solid red
curve).  The dashed red curve lies at the luminous end of the star-forming
galaxies RLF, suggesting that most QSO host galaxies have higher star-formation
rates than galaxies without a QSO.  This result is surprising, given that QSOs
are traditionally associated with ``red and dead" elliptical galaxies, at least
for radio-luminous QSOs (typically ``quasars")
\citep{dunlopEtal03,floydEtal04}.  If the host galaxies of all QSOs were
similar to the massive ellipticals with low star-formation rates that typically
host radio-luminous QSOs, we would expect the RLF of our QSO sample to follow
the extrapolation of the dashed green curve in Figure 2.  The rise in the RLF
coincides with the typical level of radio emission from star-forming galaxies,
suggesting that the host galaxies of the $\log[L_6({\rm W\,Hz}^{-1})]\sim22.4$
QSOs  are not ``red and dead" elliptical galaxies.  Instead, these host
galaxies may have spiral morphology, or the star formation may be the result of
galaxy mergers.  It has been suggested that active galaxies populating the
``green valley", with colors intermediate between the more typical blue or red
galaxy colors, may be high-Eddington-ratio ellipticals with some star
formation, or the products of mergers between massive spheroidal galaxies and
less massive gas-rich galaxies \citep[e.g., ][]{schawinskiEtal10,
  kavirajEtal09}.  Characterizing the host galaxy properties of these QSOs will
be a crucial follow-up step in investigating this result. 

\section{CONCLUSIONS}
\label{sec:conclusions}

For the first time, our observations adequately sample the radio-quiet
population; that is, they detect nearly all optically selected QSOs in a
volume-limited sample by reaching $\log[L_6({\rm W\,Hz^{-1}})]=21.5$.  Earlier
studies did not have the sensitivity needed to study the full range of
radio-quiet QSOs by reaching faint radio flux densities ($S\ll1$\,mJy) for
optical samples with significant numbers of low-redshift ($z<0.5$) QSOs to
probe the low end of the RLF.  Analyses based on radio-selected samples are
inherently biased toward the radio-loud population, but previous studies
reporting smooth luminosity distributions did not extend sufficiently faint to
study the radio-quiet population. 

The 6 GHz RLF of low-redshift color-selected QSOs is constrained by EVLA
detections of sources stronger than $20\,\mu$Jy and by the very small fraction
(6/179) of nondetections in the SDSS sample.  The strong sources constrain the
AGN contribution above $\log[L_6({\rm W\,Hz}^{-1})]\approx23.5$.  The fainter
sources imply a second radio contributor narrowly peaked around $\log[L_6({\rm
    W\,Hz}^{-1})] \approx 22.4$ and confirm the two-population model of QSO
flux densities found by \citet{kellermannEtal89}. Using the RLFs of nearby
galaxies as examples, we suggest that this second contributor is radio emission
produced by strong star formation in the QSO host galaxies, most of which
therefore cannot be ``red and dead'' massive ellipticals. 

In this Letter, we have presented the preliminary results of our investigation
to characterize the radio population of QSOs using an optically selected, volume
limited sample. We suggest that the RLF is a superposition of radio emission from
AGNs and from host star formation. It will be important to better understand the
relation between QSO emission in different wavelength regimes by characterizing the
host galaxies of these QSOs using optical and infrared colors and morphology, as
well as the radio morphology. The full details of this study will be presented in a
separate paper.

\acknowledgements

The authors would like to acknowledge helpful comments from two referees, which
have improved the clarity of the paper.  The National Radio Astronomy
Observatory is a facility of the National Science Foundation operated under
cooperative agreement by Associated Universities, Inc.  Funding for the SDSS
and SDSS-II has been provided by the Alfred P. Sloan Foundation, the
Participating Institutions, the National Science Foundation, the
U.S. Department of Energy, the National Aeronautics and Space Administration,
the Japanese Monbukagakusho, the Max Planck Society, and the Higher Education
Funding Council for England. The SDSS Web Site is http://www.sdss.org/.

\end{document}